\documentclass[aps,prd,twocolumn,floatfix]{revtex4} % {revtex4-1} % twocolumn, nofootinbib, floatfix, eqsecnum, showpacs,

\usepackage{graphics}
\usepackage{graphicx}

\usepackage{latexsym}
\usepackage[cp866]{inputenc}
\usepackage[english,russian]{babel}

\input epsf

\begin{document}

\title{\boldmath
Proposal to look for the anomalous isotopic symmetry breaking in
central diffractive production of the $f_1(1285)$ and $a^0_0(980)$
resonances at the LHC}
\author{N.~N.~Achasov and G.~N.~Shestakov}
\affiliation{Laboratory of Theoretical Physics, S.~L.~Sobolev
Institute for Mathematics, 630090 Novosibirsk, Russia}

\begin{abstract}
At very high energies, and in the central region ($x_F\simeq0$), the
double-Pomeron exchange mechanism gives the dominant contribution to
the production of hadrons with the positive $C$ parity and isospin
$I=0$. Therefore, the observation of resonances in the states with
$I=1$ will be indicative of their production or decay with the
isotopic symmetry breaking. Here, we bear in mind the cases of the
anomalous breaking of the isotopic symmetry, i.e., when the cross
section of the process breaking the isospin is not of the order of
$10^{-4}$ of the cross section of the allowed process but of the
order of $1\%$. The paper draws attention to the reactions $pp\to
p(f_1(1285)/f_1(1420))p\to p(\pi^+\pi^-\pi^0)p$ and $pp\to p(K\bar
K)p\to p(a^0_0(980))p\to p(\eta\pi^0)p$ in which a similar situation
can be realized, owing to the $K\bar K$ loop mechanisms of the
breaking of isotopic symmetry. We note that there is no visible
background in the $\pi^+\pi^-\pi^0$ and $\eta \pi^0$ channels.
Observation of the process $pp\to p(f_1(1285))p\to
p(\pi^+\pi^-\pi^0)p$ would be a confirmation of the first results
from the VES and BESIII detectors, indicating the very large isospin
breaking in the decay $f_1(1285))\to\pi^+\pi^-\pi^0$.
\end{abstract}

\maketitle

\section{Introduction} \label{Int}

The central exclusive production of hadrons, $h$, in the reactions
$pp\to p(h)p$ and $p\bar p\to p(h)\bar p$ at high energies has been
studied at the Intersecting Storage Rings and $Sp\bar pS$
accelerators at CERN and at the Tevatron collider at Fermilab and is
now being investigated in $pp\to p(h)p$ at the LHC (see, for review,
\cite{KZ07,ACF10,Kir14,GuRe14, Aal15,CMS17,FJS17}). The mass spectra
and production cross sections have been measured for a number of
hadronic systems $h$ such as $\pi\pi$ \cite{Aal15,CMS17,Ake86,
AMP87,Arm91, Bar99b,Bar99c}, $K\bar K$ \cite{Arm91,Rey98,Bar99,
Bar99a}, $\eta\pi^+ \pi^-$ \cite{Arm91a, Bar98}, $K\bar K\pi$
\cite{Bar97, Sos99}, $4\pi$ \cite{Bar97a,Bar00,Bar00a}, $\eta\pi^0$
\cite{Bar00b}, etc. Special attention was paid to the study of
resonance contributions.
%--------------------------------------------------------------------------------
\begin{figure} %[!ht]
\includegraphics[width=5.0cm]{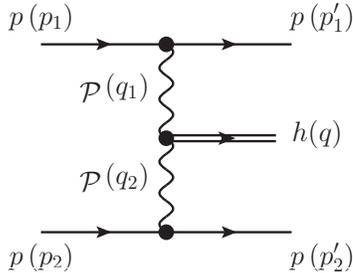}
\caption{\label{Fig20} The central production of a state $h$ by the
double-Pomeron exchange mechanism, $\mathcal{P}\mathcal{P}$, in the
reaction $pp\to p(h)p$. The 4-momenta of the initial and final
protons, $\mathcal{P}$ exchanges, and $h$ system are indicated in
parentheses; the main kinematic variables in this reaction are
$s=(p_1+p_2)^2$, $s_1=(p'_1+q)^2$, $s_2=(p'_2+q)^2$,
$M^2=q^2=(q_1+q_2)^2$, $t_1=q^2_1\simeq-\vec{q}^{\ 2}_{1\bot}$, and
$t_2=q^2_2\simeq-\vec{q}^{\ 2}_{2\bot}$.}\end{figure}
%--------------------------------------------------------------------------------

At very high energies, and in the central region, the double-Pomeron
exchange mechanism, $\mathcal{P}\mathcal{P}$, gives the dominant
contribution to production processes of hadronic resonances  with
the positive $C$ parity and isospin $I=0$ (see Fig. \ref{Fig20}).
The reaction cross sections caused by the double-Pomeron exchange
mechanism do not decrease in a power-law manner with increasing
energy \cite{KZ07,ACF10,AlGo81,GaRo80,Str86, Kai03}. Therefore,
observation of the well-known resonances in the states $h$ with
$I=1$ will be indicative of their production or decay with the
isotopic symmetry breaking. Here, we consider the examples of the
reactions $pp\to p(h)p$ in which the anomalous breaking of the
isotopic symmetry can occur.

%----------------------------------------------------------------------------------
%\vspace*{-0.2cm}
\section{\boldmath Reactions $pp\to p(f_1(1285)/f_1(1420))
p\to$ $\to p(\pi^+\pi^-\pi^0)p$} \label{SecII}

Clear signals from the $f_1(1285)$ and $f_1(1420)$ resonances with
$I^G(J^{PC})=0^+(1^{++})$ centrally produced in the reactions $pp\to
p(f_1(1285)/f_1(1420))p\to p(X^0)p$ have been observed in all their
major decay modes $X^0=\eta\pi\pi$ \cite{Arm91a,Bar98}, $K\bar K\pi$
\cite{Bar97,Sos99,GuRe14}, and $4\pi$ \cite{Bar97a,Bar00}. The
experiments were performed at incident proton laboratory momentum of
$P^p_{lab}=85$, 300, 450, and 800 GeV$/c$ or at center-of-mass
energies of $\sqrt{s}\approx 12.7$, 23.8, 29, and 40 GeV,
respectively. The data on the production cross sections of these
resonances are consistent with the $\mathcal{P}\mathcal{P}$ exchange
mechanism \cite{Arm91a,Bar98,Bar97,Sos99,Bar00,Kir14,GuRe14}. In
contrast, there is no evidence for any $0^+(0^{-+})$ contribution in
the 1.28 and 1.4 GeV regions \cite{Arm91a,Bar98,Bar97,Sos99,Bar00,
Kir14,GuRe14}. In practice, this fact can help one measure more
precisely the characteristics of the $f_1(1285)$ and $f_1(1420)$
than is possible in other experiments which see both $0^+(1^{++})$
and $0^+(0^{-+})$ states.

Thus, investigations of the $f_1(1285)$ and $f_1(1420)$ resonances
produced in central $pp$ interactions allow one to determine in a
single experiment the branching ratios for all their major decay
modes. Moreover, we pay attention that the reaction $pp\to
p(f_1(1285))p\to p(\pi^+\pi^-\pi^0)p$, due to the isotopic
neutrality of the $\mathcal{P}\mathcal{P}$ exchange mechanism, gives
a unique possibility to investigate the isospin-breaking decay
$f_1(1285)\to f_0(980)\pi^0\to\pi^+\pi^-\pi^0$ in a situation free
from any visible coherent background in the $\pi^+\pi^-\pi^0$
channel. Because of completely different experimental conditions,
such a study would be a good test of the first results from the VES
\cite{Do11} and BESIII \cite{Ab3} detectors, indicating to the
strong isospin breaking in this decay.

According to the data from the VES Collaboration \cite{Do11},
\begin{eqnarray}\label{Eq-1}
\frac{BR(f_1(1285)\to f_0(980)\pi^0\to\pi^+\pi^-\pi^0)}{
BR(f_1(1285)\to\eta\pi^+\pi^-)}\nonumber
\\ =(0.86\pm0.16\pm0.20)\%.\qquad\quad\
\end{eqnarray}
The data from the BESIII Collaboration \cite{Ab3} give
\begin{eqnarray}\label{Eq-2}
\frac{BR(f_1(1285)\to f_0(980)\pi^0\to\pi^+\pi^-\pi^0)}{
BR(f_1(1285)\to\eta\pi^+\pi^-)}\nonumber
\\ =(1.23\pm0.55)\%.\qquad\quad\qquad
\end{eqnarray}
The $f_1(1285)\to\eta\pi^+\pi^-$ decay channel has the largest
branching ratio, $BR(f_1(1285)\to\eta\pi^+\pi^-)\approx35\%$
\cite{PDG16}, among all other recorded decay channels
\cite{Arm91a,Bar98,Bar97,Sos99,Bar97a,Bar00, Kir14,GuRe14}.
Therefore, the ratios mentioned in Eqs. (\ref{Eq-1}) and
(\ref{Eq-2}) give a good reason to believe that we really deal with
the anomalously large isospin breaking in the transition
$f_1(1285)\to f_0(980)\pi^0 \to\pi^+\pi^-\pi^0$. Moreover, the
$\pi^+\pi^-$ mass spectrum observed in the decay $f_1(1285)\to
f_0(980)\pi^0\to\pi^+\pi^-\pi^0$ represents a narrow resonance
structure with a width of $10-20$ MeV located near the $K\bar K$
thresholds \cite{Do11,Ab3}. The various $K\bar K$ loop mechanisms
responsible for the decay $f_1(1285)\to f_0(980)\pi^0\to
\pi^+\pi^-\pi^0$ [i.e., the various types of transitions
$f_1(1285)\to(K^+K^-+ K^0\bar K^0)\pi^0\to f_0(980)\pi^0
\to\pi^+\pi^-\pi^0$] lead to the $\pi^+\pi^-$ mass spectrum of such
a type \cite{ADS79,AKS16,AS17,ADO15}. A significant violation of
isotopic symmetry in this transitions is a threshold phenomenon. It
occurs in the narrow region of the $\pi^+\pi^-$ invariant mass near
the $K\bar K$ thresholds due to the incomplete compensation between
the contributions of the $K^+K^-$ and $K^0\bar K^0 $ intermediate
states caused by the mass difference of the $K^+$ and $K^0$ mesons
\cite{ADS79,AKS16,AS17,ADO15,AKS15,FN1}. Certainly, the data on the
$f_1 (1285)\to f_0(980)\pi^0\to\pi^+\pi^-\pi^0$ decay need to be
clarified.

Information on the reaction $pp\to p(f_1 (1285))p \to
p(\pi^+\pi^-\pi^0)p$ could probably be extracted from the data
collected by the CERN Omega Spectrometer and Collider Detector at
Fermilab. However, for this, enthusiasts are needed, since these
facilities have long been closed. At present, the reaction $pp\to
p(f_1 (1285))p \to p(\pi^+\pi^-\pi^0)p$ could be measured, for
example, using the CMS detector at the LHC (it is interesting also
to study the related reaction $pp\to pf_1(1420)p\to
p(\pi^+\pi^-\pi^0)p$ \cite{AKS16,FN2}). Recently, the CMS
Collaboration has presented the data on the central exclusive and
semiexclusive $\pi^+\pi^-$ production in $pp$ collisions at
$\sqrt{s}=7$ TeV \cite{CMS17}. With such a huge total energy, the
energy values $\sqrt{s_1}$ and $\sqrt{ s_2}$ for the subprocesses
$p(p_1)\mathcal{P}(q_2)\to p(p'_1)h(q)$ and
$p(p_2)\mathcal{P}(q_1)\to p(p'_2)h(q)$ (see Fig. \ref{Fig20}) are
also very large. In fact, they fall into the region in which the
contributions of the secondary Regge trajectories, $\mathcal{R}$,
can be neglected in comparison with the contribution of the
$\mathcal{P}$ exchange. If we put $s_1\approx s_2$, $M\approx1$ GeV,
and $\sqrt{s}=7$ TeV and use the relation $s_1s_2\approx M^2s$
(valid for the processes in the the central region
\cite{AlGo81,GaRo80, Str86}), we find that
$\sqrt{s_1}\approx\sqrt{s_2}\approx84$ GeV. Thus, the dominance of
the $\mathcal{P}\mathcal{P}$ exchange mechanism appears to be a good
approximation at the LHC energies. Note that in the above-mentioned
experiments carried out at CERN and Tevatrov (at fixed target) the
values of $\sqrt{s_1}\approx \sqrt{s_2}$ were approximately equal to
$\approx3.6$, 4.9, 5.4, and 6.3 GeV. Therefore, in a number of
cases, when interpreting the results, it was necessary to take into
account, along with the $\mathcal{P}\mathcal{P}$ exchange mechanism,
the mechanisms involving the secondary Regge trajectories,
$\mathcal{R}$, i.e., the $\mathcal{R}\mathcal{P}$ and
$\mathcal{R}\mathcal{R}$ exchanges.

The $f_1(1285))$ resonance production with its subsequent decay to $
\pi^+\pi^-\pi^0$ can be also studied in central $pp$, $pA$, $\pi^-p
$, and $\pi^-A$ interactions at the Serpukhov acceleration in
Protvino.

%----------------------------------------------------------------------------------
%\vspace*{-0.2cm}
\section{\boldmath Reaction $pp\to p(a^0_0(980))p\to
p(\eta\pi^0)p$} \label{SecIII}

\begin{figure} %[!ht]
\includegraphics[width=6.1cm]{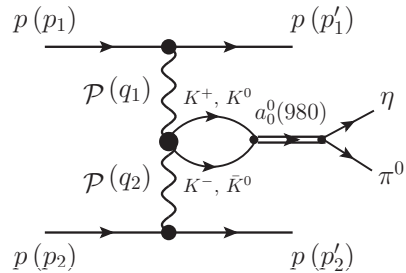}
\caption{\label{Fig21} The $K\bar K$ loop mechanism of the
$a^0_0(980)$ production in the central region via
$\mathcal{P}\mathcal{P}$ exchange.}\end{figure}

\begin{figure} %[!ht]
\includegraphics[width=5.3cm]{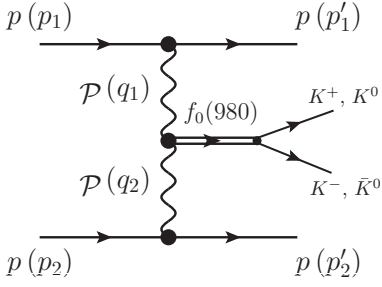}
\caption{\label{Fig22} The central production of the $f_0(980)$
resonance in the $K\bar K$ decay channels via $\mathcal{P}
\mathcal{P}$ exchange.}\end{figure}

The central production of the $a^0_0(980)$ resonance in the reaction
$pp\to p(a^0_0(980))p\to p(\eta\pi^0)p$ at the LCH energies has to
be dominated by the mechanism shown in Fig. \ref{Fig21}. The
incomplete compensation between the $K^+K^-$ and $K^0\bar K^0$
intermediate state produced in $\mathcal{P}\mathcal{P}$ collisions
leads to the isospin-breaking $a^0_0(980)$ production amplitude
which does not decrease with increasing energy. In so doing the
$\eta\pi^0$ mass spectrum has to be a narrow resonance peak (with a
width of $10-20$ MeV) located near the $K\bar K$ thresholds
\cite{ADS79,AKS16,AS17} (see, for example, Fig. \ref{Fig4} below).
Judging from the existing data, the transition amplitudes
$\mathcal{P}\mathcal{P}\to K\bar K$, that generate the process shown
in Fig. \ref{Fig21}, are dominated by the $f_0(980)$ resonance
production \cite{Kir14,GuRe14,Ake86,AMP87, Arm91,Bar99b,Bar99c,
Rey98,Bar99,Bar99a}, $\mathcal{P}\mathcal{P}\to f_0(980)\to K\bar K$
(see Fig. \ref{Fig22}).
%--------------------------------------------------------------------------------
%--------------------------------------------------------------------------------
%--------------------------------------------------------------------------------
Really, the $K^+K^-$ and $K^0\bar K^0$ mass spectra reveal powerful
enhancements near their thresholds \cite{Rey98, Bar99,Bar99a}. The
$f_0(980)$ resonance manifests itself also clearly in the
$\pi^+\pi^-$ and $\pi^0\pi^0$ mass spectra, but in the form of the
sharp dip due to the destructive interference with the large and
smooth coherent background \cite{Kir14,Ake86,AMP87,Arm91,Bar99b,
Bar99c,Aal15, CMS17}. Thus, the $a^0_0(980)$ production in the
$\eta\pi^0$ channel can predominantly occur via the
$a^0_0(980)-f_0(980)$ mixing \cite{ADS79}, i.e., owing to the $K\bar
K$ loop transition $\mathcal{P}\mathcal{P}\to
f_0(980)\to(K^+K^-+K^0\bar K^0)\to a^0_0(980)\to\eta\pi^0$. The
corresponding cross section as a function of the $\eta\pi^0$
invariant mass, $M\equiv m$, has the form (see Fig. \ref{Fig4})
\begin{eqnarray}\label{Eq7-2-1a}
\sigma(\mathcal{P}\mathcal{P}\to f_0(980)\to(K^+K^-+K^0\bar K^0)\to
a^0_0(980)\nonumber\\ \to\eta\pi^0;m)=|C_{\mathcal{P}\mathcal{P}\to
f_0}|^2\,m\Gamma_{a^0_0\to\eta\pi^0}(m)\qquad\quad\nonumber\\
\times\left|\frac{\Pi_{a^0_0f_0}(m)}{D_{a^0_0}(m)D_{f_0}(m)-
\Pi^2_{a^0_0f_0}(m)}\right|^2\,,\qquad\quad\
\end{eqnarray} where $\Gamma_{a^0_0\to\eta\pi^0}(m)$ is the width of
the $a^0_0(980)\to\eta\pi^0$ decay, $D_{a^0_0}(m)$ and $D_{f_0}(m)$
are the inverse propagators of the $a^0_0(980)$ and $f_0(980)$,
respectively, $\Pi^2_{a^0_0f_0}(m)$ is the $f_0(980)\to(K^+
K^-+K^0\bar K^0)\to a^0_0(980)$ transition amplitude (all these
functions, together with the corresponding values of the resonance
parameters, have been written in Ref. \cite{AKS16}), and
$C_{\mathcal{P}\mathcal{P}\to f_0}$ is the $f_0(980)$ production
amplitude.

%--------------------------------------------------------------------------------
\begin{figure} %[!ht]%\vspace{-2.5mm}
\includegraphics[width=7.5cm]{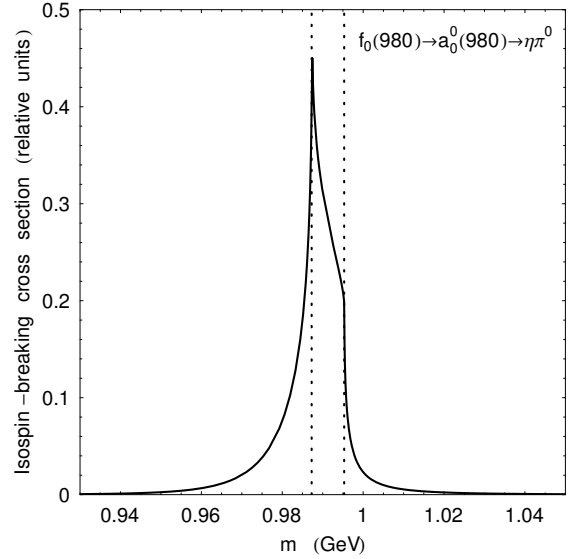}
\caption{\label{Fig4} The solid curve shows the isospin-breaking
$\eta\pi^0$ production cross section $\sigma(\mathcal{P}
\mathcal{P}\to f_0(980)\to(K^+ K^-+K^0\bar K^0)\to a^0_0(980)\to
\eta\pi^0;m)$ caused by the $a^0_0(980)-f_0(980)$ mixing and
calculated with the use of Eq. (\ref{Eq7-2-1a}). The dotted vertical
lines show the locations of the $K^+K^-$ and $K^0\bar K^0$
thresholds.} \end{figure}
%--------------------------------------------------------------------------------

Note that the $a^0_0(980)$ production cross section $\sigma(
\mathcal{P}\mathcal{P}\to(K^+K^-+K^0\bar K^0)\to a^0_0(980)
\to\eta\pi^0;m)$ can be estimated without detailing the
$\mathcal{P}\mathcal{P}\to(K^+K^-+K^0\bar K^0)$ transition mechanism
(see Fig. \ref{Fig21}), which, in principle, can be caused by not
only the $f_0(980)$ resonance contribution (see Fig. \ref{Fig22}),
but also some nonresonance $K\bar K$ production mechanism. To do
this, we use the relation valid according to the unitarity condition
near the $K\bar K$ thresholds (see Refs. \cite{AKS16, AS17} for
details)
\begin{eqnarray}\label{Eq7-2-1}
\sigma(\mathcal{P}\mathcal{P}\to(K^+K^-+K^0\bar K^0)\to
a^0_0(980)\to\eta\pi^0;m)\nonumber\\ %\qquad\qquad \ \ \
\approx|\widetilde{A}(2m_{K^+})|^2|\rho_{K^+K^-}(m)-
\rho_{K^0\bar K^0}(m)|^2\quad \nonumber\\
\times\,\frac{g^2_{a^0_0K^+K^-}}{16\pi}\,\frac{m\Gamma_{a^0_0\to
\eta\pi^0}(m)}{|D_{a^0_0}(m)|^2}\,,\qquad\quad\quad
\end{eqnarray}
where $\rho_{K\bar K}(m)=\sqrt{1-4m^2_K/m^2}\,$ at $m>2m_K$ and
$\rho_{K\bar K}(m)=i\sqrt{4m^2_K/m^2-1}\,$ at $\,0<m<2m_K$. The
resulting shape of the cross section is very similar to the solid
curve in Fig. \ref{Fig4}. The value of $|\widetilde{A}(2m_{K^+})|^2$
should be determined from the data on the $K^+K^-$ production cross
section near the threshold
\begin{eqnarray}\label{Eq7-2-2} \sigma(\mathcal{P}\mathcal{P}\to
K^+K^-;m)=\rho_{K^+K^-}(m)\,|\widetilde{A}(m)|^2\,.\ \
\end{eqnarray}
For $m$ between the $K^+K^-$ and $K^0\bar K^0$ thresholds, we get by
an order of magnitude \cite{AKS16}
\begin{eqnarray}\label{Eq7-2-3} \sigma(\mathcal{P}\mathcal{P}
\to(K^+K^-+K^0\bar K^0)\to a^0_0(980)\to \eta\pi^0;m)\nonumber\\
\approx|\widetilde{A}(2m_{K^+})|^2\times0.05.\qquad\qquad\qquad
\end{eqnarray}
The comparison of this estimate with the data on $\sigma(\mathcal{P}
\mathcal{P}\to a^0_0(980)\to\eta\pi^0;m)$ permits one to verify
their consistency with the data on $\sigma(\mathcal{P}\mathcal{P}\to
K^+K^-;m)$ and with the idea of the $K\bar K$ loop breaking of
isotopic invariance caused by the mass difference of $K^+$ and $K^0$
mesons. Note that a similar way of the checking the consistency
between the data on the decays $f_1(1285)\to\pi^+\pi^-\pi^0$ and
$f_1(1285) \to K\bar K\pi$ has been discussed in Refs.
\cite{AKS16,AS17}. Detailed formulas connecting $\sigma(\mathcal{P}
\mathcal{P}\to h;m)$ with the experimentally measured cross section
of the reaction $pp\to p(h)p$ can be found, for example, in Refs.
\cite{AlGo81, GaRo80,Str86}.

First, the central production of the $a^0_0(980)$ resonance in the
reaction $pp\to p(\eta\pi^0)p$ has been studied by the WA102
Collaboration with the use of the CERN Omega Spectrometer at
$\sqrt{s}=29$ GeV \cite{Bar00b,Sob01}. The interpretation of these
data has been discussed in Refs. \cite{CK1,AK02,AS04a,WZZ07}. Here,
we note the following. In the above experiment, the clear peaks from
$a^0_0(980 )$ and $a^0_2(1320)$ resonances have been observed in the
$\eta\pi^0$ mass spectrum. The fit \cite{Bar00b} gave the quite
usual widths of these states \cite{PDG16}: $\Gamma (a_0(980))=72\pm
16$ MeV and $\Gamma(a_2(1320)) =115\pm20$ MeV . Such a picture
indicates that at the energy $\sqrt{s_1}\approx\sqrt{s_2}\approx
\sqrt[4]{m^2_{a^0_0}s} \approx\sqrt[4]{29^2\ \mbox{GeV}^4}\approx
5.4$ GeV the secondary Regge exchanges, for which the $\eta\pi^0$
production in the central region is not forbidden by $G$ parity,
play an important role. For example, the central $a^0_0(980)$
production can proceed via $\mathcal{R}(\eta)\mathcal{R}(\pi^0)\to
a^0_0(980)$, $\mathcal{R} (a^0_2)\mathcal{R} (f_2)\to a^0_0(980)$,
and $\mathcal{R}(a^0_2 )\mathcal{P}\to a^0_0(980)$ transitions,
where the type of the secondary Regge trajectory $\mathcal{R}$ is
indicated in parentheses. At the LHC energies, the contributions of
the secondary Regge trajectories fall off appreciably and it is
natural to expect that the $a^0_0(980)$ resonance must mainly be
produced via the double-Pomeron exchange mechanism (see Fig \ref
{Fig21}), which essentially violates the isotopic invariance in the
$K\bar K$ threshold region. The narrowing of the $a^0_0(980)$ peak
in the $\eta\pi^0$ channel up to the width of $10-20$ MeV (see Fig.
\ref{Fig4}) will serve as an indicator of the changing central
production mechanism of the $a^0_0(980)$ resonance with increasing
energy.

The present work is partially supported by the Russian Foundation
for Basic Research Grant No. 16-02-00065.

%\newpage


\begin{thebibliography}{200}
\bibitem{FN1} The usual order of the isotopic symmetry breaking in
the process amplitude defined by the mass differences of the
particles in the mesonic isotopic multiplets is $\simeq(m_{K^0}-
m_{K^+})/m_{K^0}\approx 1/126$. The order of the isotopic symmetry
breaking in the process amplitude in the region between the $K^+K^-$
and $K^0\bar K^0$ thresholds due to any production mechanism of the
$K\bar K$ pairs with the definite isospin in the $S$ wave, and
without the anomalous Landau thresholds \cite{AS17,AKS16,AKS15}, in
particular, due to the $a^0_0(980)-f_0(980)$ mixing \cite{ADS79,
AS17}, is $\simeq\sqrt{ 2(m_{K^0}-m_{K^+})/m_{K^0}}\approx0.127$.
\end{thebibliography}
\end{document}